\documentclass[runningheads]{llncs}

\usepackage[caption=false]{subfig}
\usepackage{graphicx}
\usepackage{array}
\usepackage{multirow}
\usepackage{amsmath}
\usepackage{bbding}

\newcommand{\textBF}[1]{%
    \pdfliteral direct {2 Tr 0.3 w} 
     #1%
    \pdfliteral direct {0 Tr 0 w}%
}

\begin{document}
\title{Semi-Supervised Learning for Fetal Brain MRI Quality Assessment with ROI consistency}

\author{Junshen Xu\inst{1}\inst{(}\Envelope\inst{)} \and
Sayeri Lala\inst{1} \and
Borjan Gagoski \inst{2} \and
Esra Abaci Turk\inst{2} \and
\\ P. Ellen Grant\inst{2,3} \and
Polina Golland\inst{1,4} \and
Elfar Adalsteinsson\inst{1,5}
}

%
\authorrunning{J. Xu et al.}
\titlerunning{Fetal Brain MRI Quality Assessment}
%
\institute{Department of Electrical Engineering and Computer Science, MIT, \\ Cambridge, MA, USA \\\email{junshen@mit.edu} \and
Fetal-Neonatal Neuroimaging and Developmental Science Center, \\ Boston Children’s Hospital, Boston, MA, USA \and
Harvard Medical School, Boston, MA, USA \and
Computer Science and Artificial Intelligence Laboratory, MIT, \\
Cambridge, MA, USA \and
Institute for Medical Engineering and Science, MIT, Cambridge, MA, USA}

\maketitle              
\begin{abstract}
Fetal brain MRI is useful for diagnosing brain abnormalities but is challenged by fetal motion. The current protocol for T2-weighted fetal brain MRI is not robust to motion so image volumes are degraded by inter- and intra- slice motion artifacts. Besides, manual annotation for fetal MR image quality assessment are usually time-consuming. Therefore, in this work, a semi-supervised deep learning method that detects slices with artifacts during the brain volume scan is proposed. Our method is based on the mean teacher model, where we not only enforce consistency between student and teacher models on the whole image, but also adopt an ROI consistency loss to guide the network to focus on the brain region. The proposed method is evaluated on a fetal brain MR dataset with 11,223 labeled images and more than 200,000 unlabeled images. Results show that compared with supervised learning, the proposed method can improve model accuracy by about 6\% and outperform other state-of-the-art semi-supervised learning methods. The proposed method is also implemented and evaluated on an MR scanner, which demonstrates the feasibility of online image quality assessment and image reacquisition during fetal MR scans.

\end{abstract}

\section{Introduction}

Fetal brain Magnetic Resonance Imaging (MRI) is an important tool complementing Ultrasound in diagnosing fetal brain abnormalities \cite{gholipour2014fetal,malamateniou2013motion}. 
While MRI provides higher quality tissue contrast compared to Ultrasound \cite{malamateniou2013motion,kul2012contribution}, it is more vulnerable to motion artifacts because data acquisition is slow relative to the motion dynamics in the body \cite{zaitsev2015motion}. This makes it challenging to adapt MRI for fetal imaging since fetal motion is more random and larger compared to adults \cite{malamateniou2013motion}.     
The current protocol for T2-weighted fetal brain MRI attempts to mitigate motion artifacts by using time-efficient ($\sim$500 ms) readouts  per slice, such as the single-shot T2 weighted (SST2W) imaging acquisition. 
Due to safety constraints on the amount of allowable exposure to radio-frequency energy, there is a 1-2s delay between the acquisition of two consecutive slices in the stack, so obtaining an entire stack ($\sim$30 slices) takes approximately 1 minute. 
Orthogonal stacks are acquired and used to reconstruct the fetal brain volume. However, inter-slice and even intra-slice motion artifacts occur, contaminating the volume reconstruction \cite{tourbier2015efficient}. 
Therefore, in order to improve the quality, entire stacks are usually reacquired several times \cite{malamateniou2013motion,kline2014fundamental}, which is time-consuming.
Prospective detection and reacquisition of low quality slices are expected to improve both the reconstruction quality of the brain volume, as well as the efficiency of MR scans.


Several prior studies have demonstrated the potential of Convolutional Neural Networks (CNNs) for fast image quality assessment (IQA) of MRI. 
Esses et al. \cite{esses2018automated} trained a CNN for volume quality assessment of T2-weighted liver MRI. 
Sujit et al. \cite{sujit2019automated} proposed an ensemble learning method for volume quality assessment of pediatric and adult brain MRI by using multiple CNNs. 
However, several differences exist between these problems and fetal brain MR IQA. Specifically, these works aim to evaluate the quality of the entire stack of images instead of a single slice. 
Furthermore, in fetal MRI, motion is a dominant source of artifacts, typically appearing as blurs and nonuniform signal voids. Although motion is also a major source of artifacts in liver, adult brain, and cardiac MRI \cite{schreiber2017frequency}, their manifestations are different, as in this applications the  motion is more regular and smaller in range compared to the motion observed in the fetus \cite{malamateniou2013motion}. 

Since labeling large-scale medical image datasets is usually difficult and time-consuming, numerous semi-supervised learning methods have been proposed to leverage information in unlabeled data to improve the performance and robustness of deep neural networks. One general technique of semi-supervised learning is to infer pseudo labels from partially labeled data, such as self-training \cite{yarowsky1995unsupervised} and label propagation methods \cite{iscen2019label}. To yield better pseudo labels, recent methods use an ensemble of multiple neural networks which is known as self-ensembling, including temporal ensembling \cite{laine2016temporal} and mean teacher \cite{tarvainen2017mean}. In temporal ensembling, for each sample, the exponential moving average of classification outputs at different training epochs are computed and used as pseudo labels. The mean squared error (MSE) between model predictions and the pseudo labels is used as a consistency loss. One drawback of the temporal ensembling method is that it needs to keep track of the pseudo labels which is memory-consuming for large datasets. To address this problem, mean teacher method is proposed, which instead of using an ensemble of network outputs, aggregates the parameters of networks at different training step to build a teacher model. The system consists of two models with the same architecture, i.e., student and teacher. The student model is updated with gradient during training, while the teacher model is the exponential moving average of the student model. The prediction of the teacher model is considered as pseudo label, and a consistency loss similar to temporal ensembling, is enforced between the predictions of student and teacher models. The consistency loss in self-ensembling method can also be interpreted as a regularization that smooths the network around unlabeled data. Following this interpretation, Miyato et al. proposed virtual adversarial training (VAT) \cite{miyato2018virtual} where they enforced consistency between predictions of original images and corresponding adversarial samples. These semi-supervised methods have also found their way into application of medical imaging, such as nuclei classification \cite{su2019local} and gastric diseases diagnosis \cite{shang2019leveraging}.

In this work, we proposed a novel semi-supervised learning method for fetal MRI quality assessment. Our method extends the mean teacher model by introducing a region-of-interest (ROI) consistency for fetal brain, which let the network focus on the fetal brain ROI during feature extraction, and thus improves the accuracy of detecting non-diagnostic MR images. Evaluation showed that our method outperformed other state-of-the-art semi-supervised methods. We also implemented and evaluated the proposed method on a MR scanner, demonstrating the feasibility of online image quality assessment and image reacquisition during fetal MR scans.

\section{Methods}

\subsection{Mean Teacher Model}

In semi-supervised learning, let $\{x_1,x_2,...,x_{N_l}\}$ be the labeled dataset with labels $\{y_1,y_2,...,x_{N_l}\}$ and let $\{x_{N_l+1},x_{N_l+2},...,x_{N}\}$ be the unlabeled dataset. The mean teacher model \cite{tarvainen2017mean} consists of two networks with the same architecture, i.e., student network and teacher network, whose parameters are denoted as $\theta$ and $\theta'$ respectively. 

During training, the student network is updated by minimizing the following loss function:
\begin{equation}
\begin{split}
L_{\text{MT}}&=L_{\text{cls}}+\lambda L_{\text{con}}\\
&=\frac{1}{N}\sum_{i=1}^{N_l} H(y_i, f_\theta(x_i,\eta)) + \frac{\lambda}{N}\sum_{i=1}^{N} D_{\text{KL}} (f_{\theta'}(x_i,\eta')||f_\theta(x_i,\eta))\\
\end{split}
\end{equation}
The first term is the classification loss for labeled data, which is the cross entropy between student network prediction $f_\theta(x_i,\eta)$ and label $y_i$. The second term is the consistency loss between predictions of student and teacher networks. Inspired by VAT \cite{miyato2018virtual}, we use Kullback–Leibler (KL) divergence to measure the distance between the student and teacher predictions, instead of MSE as used in the original mean teacher method \cite{tarvainen2017mean}, where $\eta$ and $\eta'$ denote the noise perturbation for the two networks and $\lambda$ is the weight of consistency loss. The teacher network is updated as follows: $\theta'_{t+1}=\alpha\theta'_t+(1-\alpha)\theta_t$, where $\alpha$ is the coefficient and $t$ is training step.

\subsection{Brain ROI consistency}

\begin{figure}
    \centering
    \includegraphics[width=\textwidth]{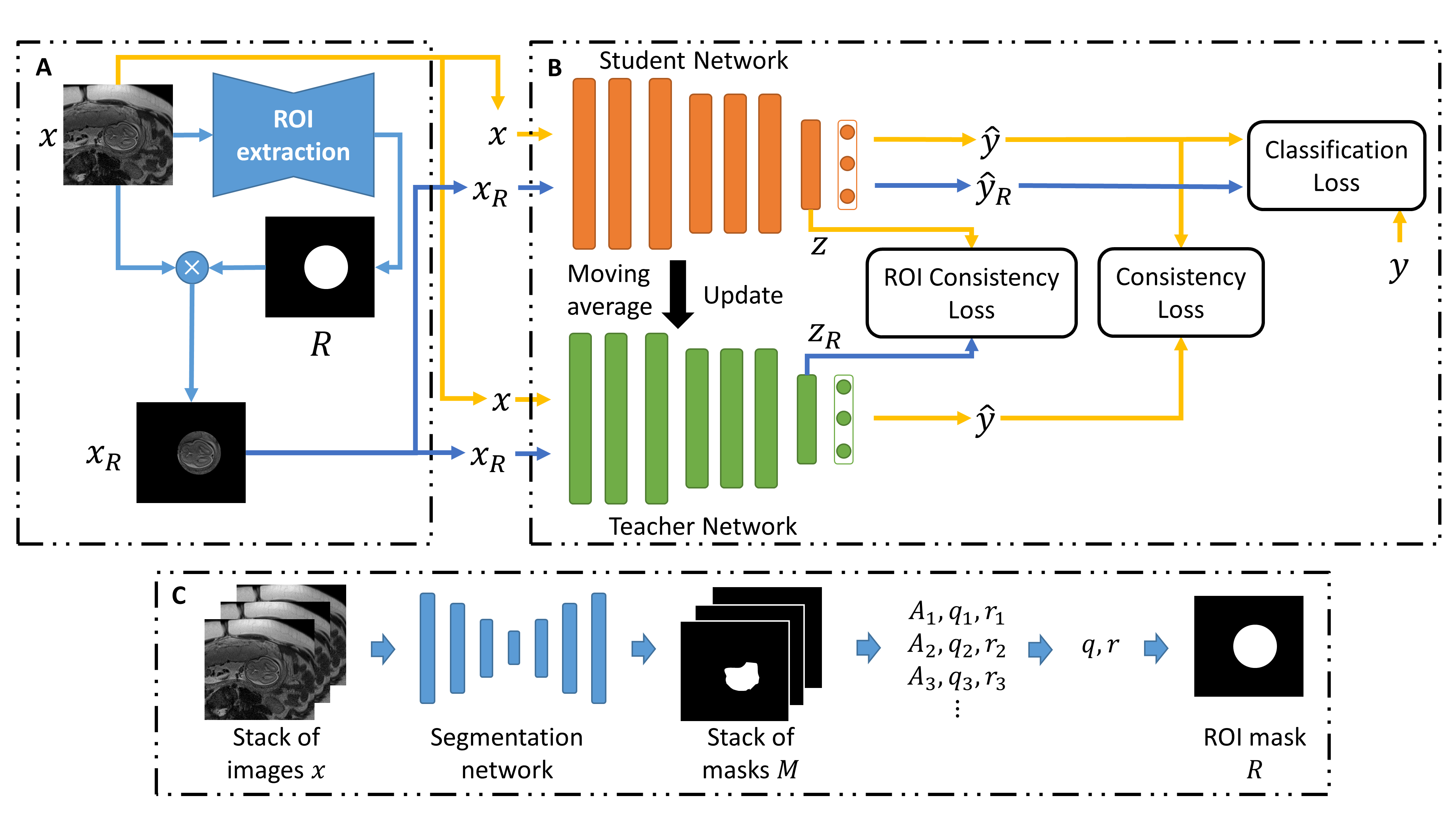}
    \caption{Overview of the proposed method. A) Brain ROI extraction. B) Mean teacher model with ROI consistency loss.  C) Details of ROI extraction algorithm.}
    \label{fig:method}
\end{figure}

In fetal brain MRI, the brain occupies a small portion of the image due to imaging parameter constraints \cite{gholipour2014fetal}. However, the fetal brain is the ROI relevant for fetal brain MRI IQA since only the artifacts occurring in the brain affect diagnostic quality of the image. Therefore, it is essential to train the model to focus on features within the brain ROI. To fulfill this goal, We propose an ROI consistency loss to regularize the network. The overall architecture of the proposed mean teacher model with brain ROI consistency is shown in Fig. \ref{fig:method}. 

First, we introduce an ROI extraction module (Fig. \ref{fig:method}A). For each image $x$, it produces a brain ROI mask $R$. $x_R=x\odot R$ is the masked image, where $\odot$ is the Hadamard product. The implementation of ROI extraction relies on a segmentation model. We utilize a trained U-Net in \cite{salehi2018real} to segment fetal brains from MR slices. However, since the segmentation network is trained on images with different acquisition parameters, it may yield inaccurate segmentation masks and fail to detect the brain ROI for some slices in our dataset. To improve robustness of ROI detection, instead of using the output of segmentation network directly, we aggregate the masks of images belonging to the same scan to generate a single ROI mask for the whole stack of images. The proposed algorithm is described in Fig. \ref{fig:method}C. A stack of images are fed into the pretrained network to generate raw masks. For each mask $M_i$ in the stack, its area $A_i$, center $q_i$ and radius $r_i$ are computed. We exclude those masks with area less than a threshold $A_{\min}$, which are assumed to be inaccurate, and let $B=\{i| 1\le i\le S, A_i\ge A_{\min}\}$ be the set of remaining slices. We then compute the area-weighted mean and variance of the centers over $B$, i.e., $q = \frac{1}{|B|}\sum_{i\in B} A_i q_i$ and $\sigma^2 = \frac{1}{|B|}\sum_{i\in B} A_i ||q_i - q||_2^2$. The final ROI mask $R$ is defined as the circle centered at $q$ with radius $r=\sigma + \max_{i\in B}{r_i}$.

The goal of ROI consistency loss is to make the network focus on brain ROI. Let $z$ be the output feature of the last convolution layer. $z_{\theta'}(x_i\odot R_i, \eta)$ is the feature of ROI extracted by the teacher network and $z_{\theta}(x_i, \eta)$ is the feature of the original image extracted by the student network. We want these features to be close to each other, so that the student can learn to detect the brain ROI from the whole image. The ROI consistency loss are defined as the MSE between these two features:
\begin{equation}
    L_{\text{con-roi}}=\frac{1}{N}\sum_{i=1}^{N}||z_{\theta'}(x_i\odot R_i, \eta) - z_{\theta}(x_i, \eta)||_2^2
\end{equation}
The ROI consistency loss use the feature of masked images extracted by the teacher network as reference. To guide the teacher network to learn meaningful features from the masked images, the classification loss for masked images in the labeled dataset is used as a regularization which is denoted as $L_\text{cls-roi}$.
\begin{equation}
    L_{\text{cls-roi}}=\frac{1}{N}\sum_{i=1}^{N_l} H(y_i, f_\theta(x_i\odot R_i,\eta))
\end{equation}

We also adopted conditional entropy as an additional loss:
\begin{equation}
    L_{\text{ent}}=H(y|x)=\frac{1}{N}\sum_{i=1}^{N} H(f_\theta(x_i,\eta),  f_\theta(x_i,\eta))
\end{equation}
which is able to exaggerate the prediction of the network on each data point \cite{miyato2018virtual}. Therefore, the total loss of the proposed method is as follows.
\begin{equation}
    L=L_{\text{cls}}+L_\text{cls-roi}+\lambda L_{\text{con}}+\beta L_{\text{con-roi}}+\gamma L_{\text{ent}}
\end{equation}
where $\lambda$, $\beta$ and $\gamma$ are weight coefficients. 

At the first couple of epochs, the teacher network cannot provide a reliable guide to the student network. For this reason, we use a ramp-up function $w(t) = \exp[-5(1 - \min(t, T) / T)^2]$ for coefficients $\lambda, \beta$ and $\gamma$, where $t$ is the current epoch and $T=5$.

\section{Experiments and Results}

\subsection{Dataset}

A total of 217129 images were obtained from 644 previously acquired research and clinical scans of mothers with singleton pregnancies and no pathologies, ranging in gestational age between 19 to 37 weeks.  Scans were conducted at Boston Children's Hospital with Institutional Review Board approval. Scans were acquired using the SST2W sequence with median echo time $\text{TE}=115\text{ ms}$, repetition time $\text{TR}=1.6\text{ s}$, field of view 31 cm, and voxel size of $1.2\times1.2\times3\text{ mm}^3$.

\begin{figure}
    \centering
    \includegraphics[width=0.7\textwidth]{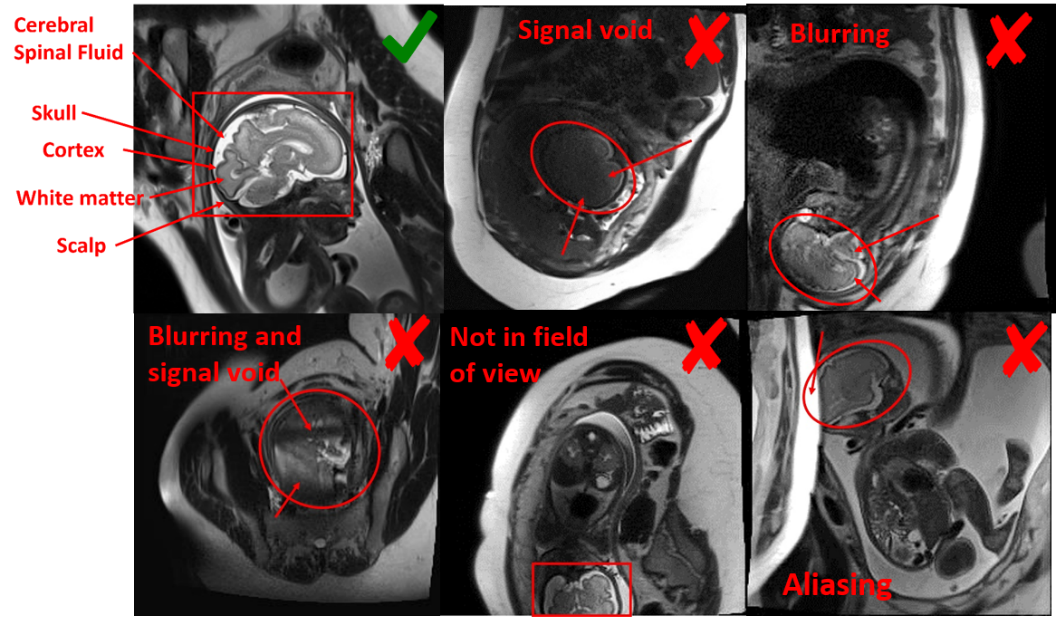}
    \caption{Representative examples of diagnostic and nondiagnostic quality fetal brain MRI}
    \label{fig:haste}
\end{figure}

A set of 11223 images from 42 subjects are selected as labeled set and classified into three categories: diagnostic (D), non-diagnostic (N) and images without brain region of interest (W). Diagnostic images were characterized by sharp brain boundaries while non-diagnostic images were characterized by artifacts that occlude such features (Fig. \ref{fig:haste}). Motion artifacts manifest as signal void and blurring over the brain region. Other artifacts manifest as aliasing or the fetus not being in the field of view. A research assistant trained under radiologists labeled the dataset. The labeled dataset is divided into training (7717 images), validation (1782 images), and test (1724 images) set, where the test set consists of subjects different from training and validation sets. 

\subsection{Experiments Setup}

We adopted ResNet-34 \cite{He_2016_CVPR} as the backbone for student and teacher networks and set $\alpha=0.994$ and $\lambda=\beta=\gamma=1$, unless otherwise stated.

To evaluate the proposed method, we compare it with other three methods, including supervised learning, mean teacher (MT) \cite{tarvainen2017mean} and virtual adversarial training (VAT) \cite{miyato2018virtual}. In addition to accuracy, we also adopted area under the ROC curves (AUC) for non-diagnostic images (N) as performance metric, since in clinical practice we are interested in detecting non-diagnostic slices and reacquiring them during MR scan.

For each method, we train the model using 1000, 2000, 4000 and all labeled data in training set (7717). For semi-supervised method, all unlabeled data are used for training. We used a batch size of 384. To balance the number of labeled and unlabeled data seen by the model, in each batch, 96 images are drawn from labeled dataset while the remains are unlabeled data. We run each experiment for 5 times and report the mean and standard deviation of evaluation metrics.

All Neural networks were implemented with PyTorch and trained on a server with an Intel Xeon E5-1650 CPU, 128GB RAM and four NVIDIA TITAN X GPUs. Adam \cite{kingma2014adam} optimizer is used with an initial learning rate of $5\times10^{-3}$, and cosine learning rate decay. 

\subsection{Results}

Results of accuracy and AUC are reported in Table \ref{tab:metric}. Results show that, the proposed method outperforms other state-of-the-art semi-supervised learning method in terms of both accuracy and AUC of non-diagnostic image. Additionally, comparing with supervised learning, the proposed approach increases accuracy and AUC by 5.82\% and 0.084 respectively by learning extra information of large scale unlabeled dataset. We can also see that for smaller labeled training set (e.g., 1000 labels) the gain in accuracy from unlabeled data is higher. Besides, ablation studies were performed by setting $\lambda$, $\beta$ or $\gamma$ to zero to evaluate the contribution for each regularization. Results show that all the three regularization terms in our method can improve the performance of network. 

\begin{table}[!ht]
    \caption{Accuracy $\pm$ std (\%) and AUC for non-diagnostic image $\pm$ std over 5 runs.}
    \label{tab:metric}
    \begin{center}
    \begin{tabular}{c|c|cccc}
    \hline
    metric & method     & 1000 labels& 2000 labels & 4000 labels & all labels\\
    \hline
    \multirow{7}{*}{Acc.} & supervised & $75.58\pm0.93$ & $77.40\pm0.68$ & $79.33\pm0.94$ & $79.37\pm0.38$ \\
    & VAT \cite{miyato2018virtual} & $76.06\pm2.18$ & $77.51\pm1.90$ & $80.45\pm2.35$ & $81.25\pm1.21$ \\
    & MT \cite{tarvainen2017mean} & $79.27\pm0.85$ & $80.35\pm0.56$ & $81.21\pm0.81$ & $81.89\pm0.63$ \\
    \cline{2-6}
    & proposed & $\textBF{82.87\pm0.92}$ & $\textBF{83.73\pm0.86}$ & $\textBF{84.37\pm0.37}$ & $\textBF{85.19\pm0.19}$ \\
    & $\lambda=0$ & $80.88\pm1.07$ & $81.38\pm0.70$ & $82.47\pm0.42$ & $82.88\pm0.31$ \\
    & $\beta=0$& $80.78\pm0.66$ & $82.01\pm1.03$ & $83.27\pm0.34$ & $83.81\pm0.52$ \\
    & $\gamma=0$ & $80.61\pm0.26$ & $80.92\pm0.61$ & $82.68\pm0.53$ & $83.77\pm0.40$ \\
    \hline
    \multirow{7}{*}{AUC} & supervised & $0.788\pm0.016$ & $0.818\pm0.012$ & $0.826\pm0.008$ & $0.815\pm0.012$ \\
    & VAT \cite{miyato2018virtual} & $0.815\pm0.021$ & $0.822\pm0.014$ & $0.833\pm0.017$ & $0.844\pm0.044$ \\
    & MT \cite{tarvainen2017mean} & $0.831\pm0.008$ & $0.851\pm0.005$ & $0.856\pm0.011$ & $0.864\pm0.006$ \\
    \cline{2-6}
    & proposed & $\textBF{0.869\pm0.008}$ & $\textBF{0.881\pm0.003}$ & $\textBF{0.889\pm0.007}$ & $\textBF{0.899\pm0.006}$ \\
    & $\lambda=0$& $0.829\pm0.007$ & $0.822\pm0.001$ & $0.841\pm0.011$ & $0.854\pm0.008$ \\
    & $\beta=0$ & $0.854\pm0.006$ & $0.872\pm0.005$ & $0.875\pm0.004$ & $0.887\pm0.006$ \\
    & $\gamma=0$ & $0.855\pm0.009$ & $0.860\pm0.003$ & $0.878\pm0.006$ & $0.882\pm0.005$ \\
    \hline
    \end{tabular}
    \end{center}
\end{table}


\subsection{Online Implementation}

\begin{figure}[ht]
\centering
\subfloat[\label{fig:simulate}]{%
  \includegraphics[width=0.34\textwidth]{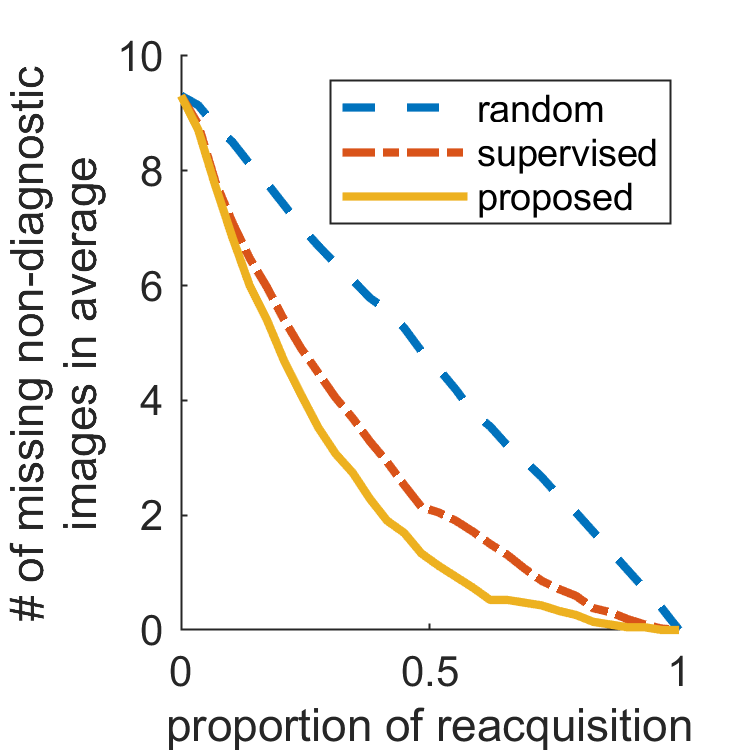}%
}\hfil
\subfloat[\label{fig:online}]{%
  \includegraphics[width=0.65\textwidth]{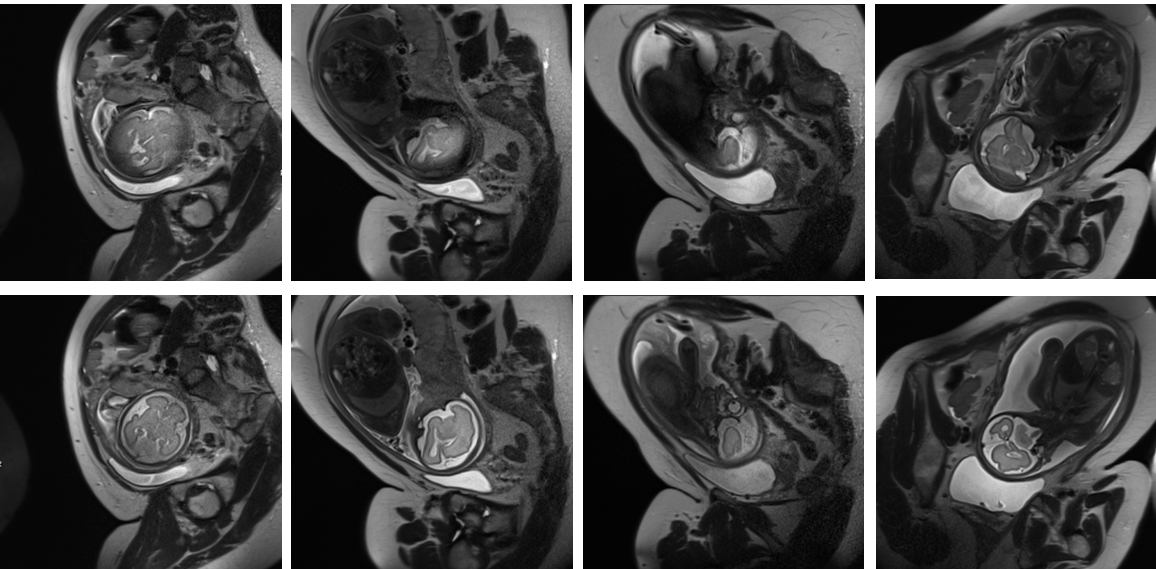}%
}
\caption{(a) Number of non-diagnostic slices that are not detected by the IQA pipeline. (b) Four examples from three separate in vivo scans showing motion artifacts in the originally acquired images (top row), and much cleaner images when the same slice locations were re-acquired (bottom row).}
\label{fig:simluate_online}
\end{figure}

To further evaluation the proposed method and its performance in clinical practice, we developed and implemented a pipeline that runs the IQA CNN during fetal MR scans to assign a IQA score to each slice and reacquire those slices with low IQA scores. The trained CNN is deployed on a GPU (NVIDIA 1050Ti) equipped computer which is connected to the scanner’s internal network. In each scan, $N_{acq}$ slices were acquired and the IQA scores are computed as $s=1-P_N$, where $P_N$ is probability of non-diagnostic image. Then the $N_{re}$ slices with lowest IQA scores were reacquired. The proportion of re-acquisition is denoted as $q=N_{re}/N_{acq}$. We performed a simulation study on the test set consisting of stacks of images with 20 to 40 slices where about one third of the images are of low quality in average (in the worst case, over 60\% of slices in a stack are contaminated by motion artifacts). The number of missing non-diagnostic images is shown in Fig. \ref{fig:simulate}, where 'random' means random re-acquisition. The proposed method outperforms the supervised baseline and only misses one non-diagnostic slice in average when $q=50\%$.

For in vivo study, fetal scans were performed on a 3T MR scanner with $N_{acq}=20, q=0.5$. Fig. \ref{fig:online} shows 4 images from 3 separate scans, where the originally acquired slices (top row) were motion degraded, and the re-acquired ones (bottom row) were not. These results demonstrated the feasibility of online detection of non-diagnostic MR images during fetal scans using the proposed deep learning method.


\section{Conclusions}
In this paper, we proposed a novel semi-supervised learning method for fetal MRI quality assessment. Our method extend the mean teacher model by introducing a ROI consistency for fetal brain which let the network focus on brain ROI during feature extraction and therefore improve the accuracy of detecting non-diagnostic MR images. Evaluation showed that our method outperformed other state-of-the-art semi-supervised methods as well. We also implemented and evaluated the proposed method on a MR scanner, demonstrating the feasibility of online image quality assessment and image requisition during fetal MR scans, which can work in tandem with fetal motion tracking algorithm \cite{xu2019fetal} to improve image quality as well as efficiency of imaging workflow.



\bibliographystyle{splncs04}






\end{document}